# Synthesis and investigation of the properties of organic-inorganic perovskite films with non-contact optical methods

V.P. Kostylyov[1], A.V. Sachenko[1], V.M. Vlasiuk[1], I.O. Sokolovskyi[1],
S.D. Kobylianska[2], P.V. Torchyniuk[2], O.I. V'yunov[2], A.G. Belous[2]

[1]V. Lashkaryov Institute of Semiconductor Physics, National Academy of Sciences of Ukraine, 41 pr. Nauki, Kiev 03028 Ukraine, e-mail: vkost@isp.kiev.ua
[2]Vernadsky Institute of General and Inorganic Chemistry, National Academy of Sciences of Ukraine, 32/34 Prospect Palladina, Kiev 03142,Ukraine

Abstract

Presented in this work are the results of our study of the photoelectric properties of perovskite $CH_3NH_3PbI_{2.98}Cl_{0.02}$ films deposited on a glass substrate using the spin-coating method. The unit cell parameters of the perovskite are determined using x-ray diffractometry. It is shown that the film morphology represents a net of non-oriented needle-like structures with significant roughness and porosity.

In order to investigate the properties of the films obtained, non-contact methods were used, such as transmission and reflection measurements and the measurements of the spectral characteristics of the small-signal surface photovoltage.

The method of spectral characteristics of the low-signal surface photovoltage and the transmission method reveal information about the external quantum yield in the films studied and about the diffusion length of minority carriers in the perovskite films. As a result of this analysis it has been established that the films obtained are naturally textured, and their bandgap is 1.59 eV. It is shown that in order to correctly determine absorption coefficient and the bandgap values, Urbach effect should be accounted for. Minority carriers' diffusion length is longer than the film thickness, which is equal 400 nm. The films obtained are promising materials for solar cells.

# Introduction

An increase of the world energy consumption with hydrocarbons and atomic energy as primary sources has created a number of ecological, technological, social, and recently also economic problems. This has led to a growing of interest in the renewable energy sources. One of the most attractive and promising renewable sources of energy is photovoltaics, that is, direct transformation of solar radiation energy into electricity. Increasing the efficiency of photovoltaic energy conversion and decreasing the production cost per unit of energy remains an important problem since quite a long time.

The promising candidates that can solve the above-pointed problems are solar cells (SCs) based on the organic-inorganic chalcogenides of metals with perovskite structure [1,2]. These are direct-bandgap materials, implying high optical absorption coefficient value. SCs based on these materials belong to the second generation of thin-film SCs [3]. As compared to the traditional SCs based on monocrystalline silicon, their fabrication technology is simpler, because it does not require high temperatures, which reduces the price of these SCs. Besides, during the several years of their investigation, the efficiency of these elements has increased to 20,9% for the SC with the area of 1 $cm^2$ and to 22.7% for the area of 0.09 $cm^2$ [4,5].

The most common methods to fabricate organic-inorganic metal chalcogenides are the so-called one-step solution deposition processes, in which all components are first solved in an organic solvent, such as N-dimethylformamide (DMF), then deposited on a substrate, and finally are thermally treated [6,7]. Production of perovskite thin films is possible using many different deposition methods, such as spin-coating [8], deposition by dipping into a solution [9], blade-coating [10], sputtering [11], and vacuum deposition [12]. Also, the use of other solvents has been reported, such as butyrolactone [13] or dimethyl sulfoxide [14]. All the factors (deposition methods and solvents used) have a strong effect on the crystallization processes, therefore the films obtained are characterized by different morphologies (the shape and the size of the grains), and, correspondingly, by different crystal structure defects both on the grain boundaries and in the films. In polycrystalline films, grain size, their defect types and orientation strongly affect their optical and photoelectric properties.

For photoconversion, optical and photoelectric characteristics are the most crucial. As a rule, they are studied in the already fabricated SCs, which consist of several layers in addition to the perovskite layer. As a result, the characteristics measured are integrated, that is, it is often difficult to single out the contributions of individual layers, which is important when production technologies are being developed.

In this work, it is proposed to use the method of spectral characterization of the small-signal surface photovoltage together with the measurements of the optical properties (transmission and reflection) for characterization and development of fabrication technologies of perovskite films of different microstructures. By definition, the small-signal surface voltage $V_{ph}$ is always much smaller than the thermal voltage, $V_{ph} << kT/q$, where $k$ is Boltzmann's constant, $T$ is the temperature, and $q$ is the elementary charge. At $T$=300 K, $kT/q$ = 25.9 mV.

The methods proposed allow one to establish the mechanism of short-circuit current formation, to obtain the bandgap value in a perovskite film as well as the diffusion length and the lifetime (if the diffusion coefficient is known) of the non-equilibrium minority charge carriers, and to estimate surface recombination velocity from the short-wavelength part of the photosensitivity spectrum.

**Experimental methods**

*Synthesis method.* Lead iodide $PbI_2$, methylammonium chloride $CH_3NH_3Cl$ (C.P.), and methylammonium iodide $CH_3NH_3I$ that was pre-synthesized were used as the input reagents [15]. In order to stabilize the perovskite structure, partial replacement of iodine with chlorine was carried out [16] using methylammonium chloride $CH_3NH_3Cl$ (C.P.). Dried dimethylformamide (DMF, C.P.) was used as a solvent.

To obtain the $CH_3NH_3PbI_{2.98}Cl_{0.02}$ films, the input reagents $PbI_2$, $CH_3NH_3I$, and $CH_3NH_3Cl$ were dissolved in DMF in stoichiometric proportions and mixed for 1 hour at 70°C. The synthesis was performed in a dry box. The so obtained transparent solution was then deposited onto a pre-cleaned substrate by spin-coating at 1200 rotations per minute during 30 seconds. Glass or ITO-coated glass (denoted hereafter as ITO/glass) were used as substrates. Thermal treatment of the films was carried out on a pre-heated stove at 90°C for 30 minutes.

The products were characterized by x-ray powder diffraction taken on the apparatus DRON-4-07 (CuKα-radiation, 40 кV, 18 mA) in the 2θ=10-120° range with the step size of 0,02° and exposition time of 6 s. The unit cell parameters were determined using the Rietveld full-profile analysis method of the data.

***Measurements of the physical characteristics.*** The spectral characteristics of the surface photovoltage were measured in the wavelength range $\Delta\lambda = 400\text{-}900$ nm on the perovskite $CH_3NH_3PbI_{2.98}Cl_{0.02}$ films deposited on glass with an ITO layer. The measurements were performed at a constant flux of photons of monochromatic light, or constant irradiance. The spectra obtained from these measurements are proportional to the external and internal quantum efficiency. Surface photovoltage measurements were performed with a non-destructive method using a press-on ITO electrode with the area ~7×7 mm² deposited on mica of ~5 μm thickness (Fig. 1). The spectral measurements were carried out on a set-up for the determination of spectral characteristics of photoconverters in the Center for testing of photoconverters and photoelectric batteries at the V.E. Lashkaryov Institute of Semiconductor Physics of the NAS of Ukraine.

The transmission spectra in the wavelength range $\Delta\lambda = 400\text{-}900$ nm were measured on the perovskite $CH_3NH_3PbI_{2.98}Cl_{0.02}$ film samples deposited on glass without an ITO layer. As a photodetector, a silicon photodiode was used.

The reflection coefficient was estimated for the wavelength of 632.8 nm. The measurements have revealed that reflection had diffusive character, and the reflection coefficient was quite small, estimated to be ~5%.

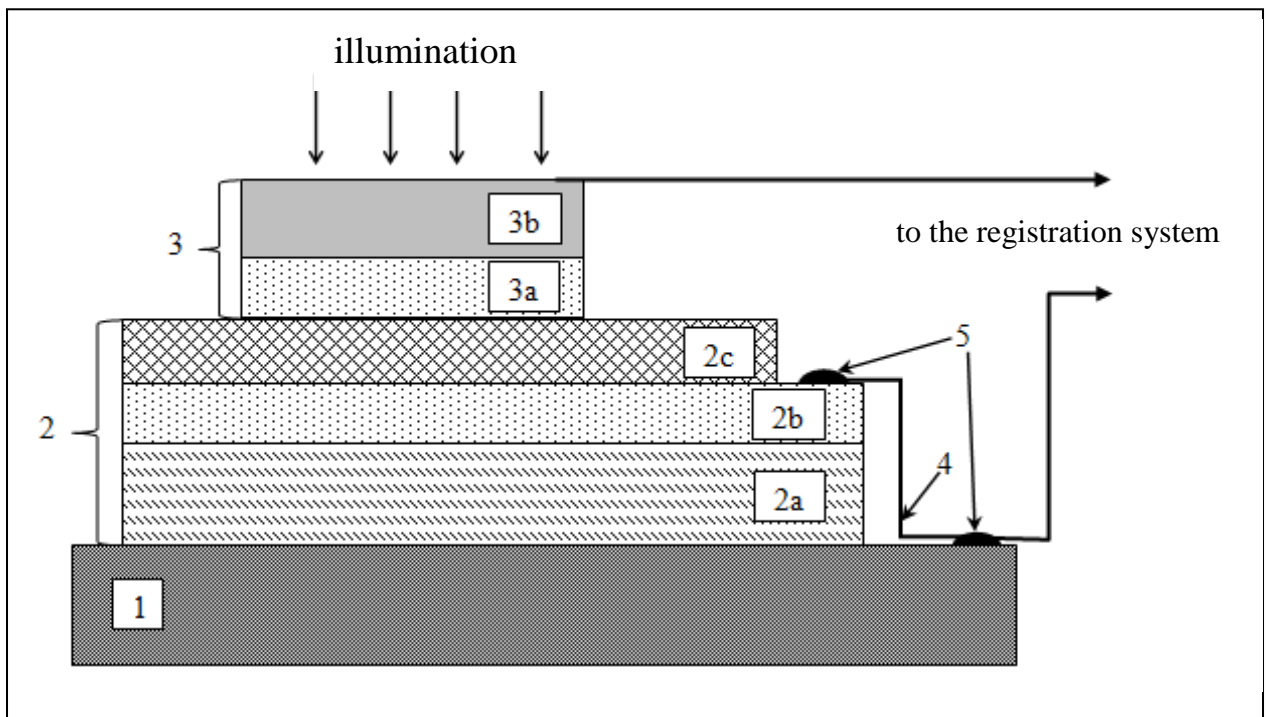

Fig. 1. Schematics of the experimental set-up for the measurements of the surface photovoltage: 1-contact table, 2-the sample under study, 2a-glass substrate, 2b-ITO, 2c-perovskite film, 3-transparent press-on electrode, 3a-mica, 3b-ITO, 4-copper conductor, 5-Ohmic contacts.

**Results and discussion**

To determine the unit cell parameters of the $CH_3NH_3PbI_{2.98}Cl_{0.02}$ material synthesized using the full-profile Rietveld method, the x-ray diffraction patterns of the single-phase samples were used, one of which is shown in Fig. 2. Calculations of the structure parameters indicate that this diffractogram corresponds to the tetragonal symmetry (*I4/mcm* space group, № 140), which agrees with the literature data [17].

For the calculations, the atomic coordinates from this work were used for Pb (4c) 000; I1(8h) xy0; I2(4a) 00¼; C (16l) xyz; N (16l) xyz. For the $CH_3NH_3PbI_{2.98}Cl_{0.02}$ film the following lattice parameters were obtained: $a = 0.8870(2)$ Å, $c = 1.2669(8)$ Å, $V = 0.9968(7)$ Å$^3$.

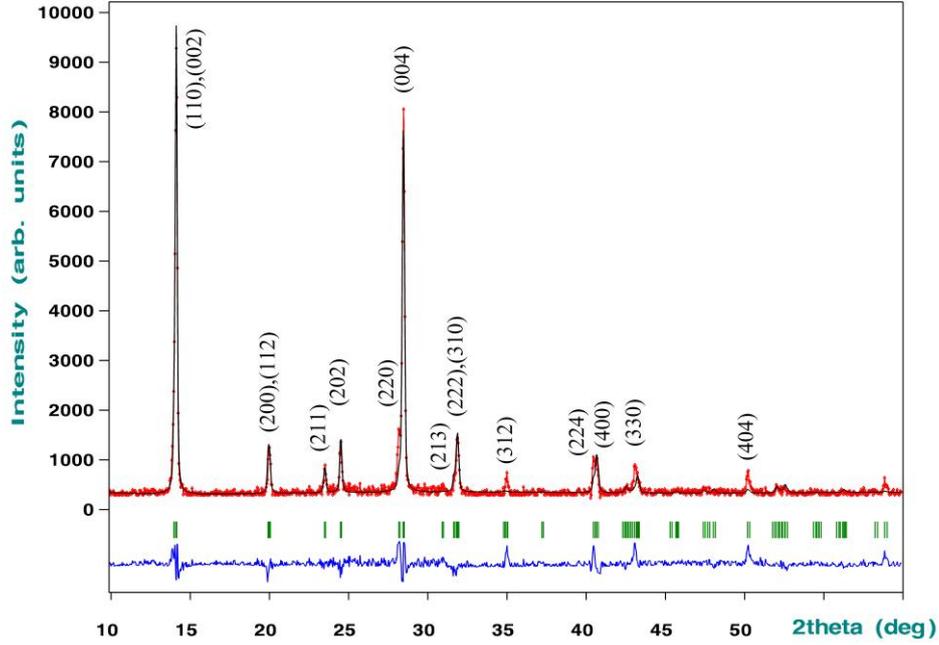

Fig. 2. Experimental (symbols) and theoretical (curves) x-ray diffractograms of a $CH_3NH_3PbI_{2.98}Cl_{0.02}$ film sample after thermal processing at 90 °C. The vertical lines indicate peak locations, with Miller indices given in the brackets. The difference curve is shown below.

The microstructure analysis of the films obtained, see Fig. 3, shows that the ITO/glass substrate is incompletely covered by $CH_3NH_3PbI_{2.98}Cl_{0.02}$. The film morphology can be described as a net composed of non-oriented needle-like structures with a broad range of length to width ratio and significant film roughness and porosity.

In general, the spectral dependence of the low-signal surface photovoltage $V_{ph}(\lambda)$ can be described as [18]

$$V_{ph}(\lambda) = A_1(\varphi_s, \lambda) EQE(\lambda) = A_2(\varphi_s, \lambda) IQE(\lambda), \qquad (1)$$

where $\varphi_s$ is the band bending on the illuminated side of the film, $\lambda$ is the light wavelength, $A_1$ and $A_2$ are the coefficients with the dimensions of Volts, $\alpha(\lambda)$ is light absorption coefficient, and $EQE(\lambda)$ and $IQE(\lambda)$ are the external and internal quantum efficiencies. The relation between the external and internal quantum efficiencies for a non-absorbing film has the form [19]

$$EQE(\lambda) = (1 - R(\lambda))IQE(\lambda), \qquad (2)$$

where $R(\lambda)$ is the reflection coefficient.

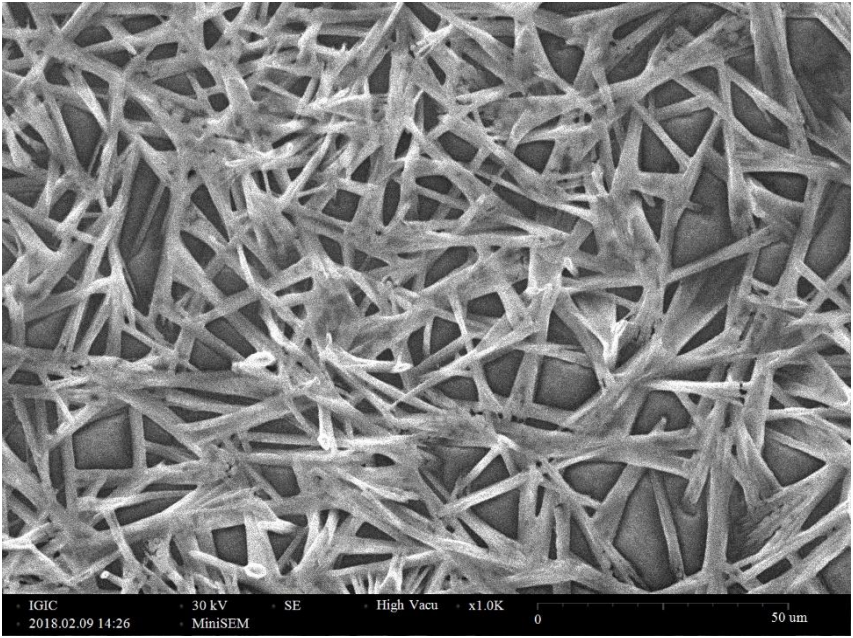

Fig. 3. $CH_3NH_3PbI_{2.98}Cl_{0.02}$ film microstructure after its thermal processing at 90 °C.

In the case relevant for photoconversion (the minority carrier diffusion length $L$ is much longer than the film thickness $d$), according to [20], the dependence of the internal quantum efficiency on $\lambda$ near absorption edge for a parallel-plane structure with a single reflection has the form

$$IQE_{nt}(\lambda) = 1 - \exp(2\alpha(\lambda)d). \qquad (3)$$

For a textured film, which has either natural or artificially created surface roughness, it is given by [20]

$$IQE_t(\lambda) = \left(1 + \left(4\alpha(\lambda)d\,n_r^2\right)^{-1}\right)^{-1}. \qquad (4)$$

Here $d$ is the perovskite film thickness and $n_r$ is its refraction coefficient.

Note that for $L \leq d/3$ the internal quantum efficiency is well described by the well-known formula [21]

$$IQE(\lambda) = \frac{\alpha(\lambda)L}{1 + \alpha(\lambda)L}, \quad (5)$$

where $L$ is the minority carrier diffusion length.

When $L \leq d/3$, the diffusion length $L$ can be found from (5).

As shown in [22], the probability of photon absorption increases due to an increase of its path length from the value of $2d$ in a plane-parallel structure with a flat surface to the value of $4n_r^2 d$ in a textured structure. This is the difference between the expressions (3) and (4). Texturing is known to be responsible for the higher values of the quantum efficiency and the short-circuit current, and hence of the photoconversion efficiency. An increase of the photoconversion efficiency is not only due to an increased photon path length, which leads to a broadening of quantum efficiency with $\lambda$ and to its shifting towards the longer wavelengths; it is also due to a reduced reflection coefficient value. This is related to the non-perpendicular light incidence on the textured elements of the structure, which leads to multiple absorption.

It was established in [23] that the absorption edge in the FAPbI$_3$-based SCs can be described by the generalized expression

$$IQE^*(\lambda) = \left(1 + \left(4\alpha(\lambda)n_r^2/b\right)^{-1}\right)^{-1}, \quad (6)$$

where $b \geq 1$ is a non-dimensional parameter equal to the ratio of the longest photon path length possible, $4n_r^2$, to its real value. This ratio depends on the texturing quality and film thickness.

In this case, the value $b = 1$ was obtained, in spite of the fact that the structure was not textured artificially. Hence, because of the non-planar arrangement of the grains that compose the perovskite, its natural texturing occurs.

We note that the case when the expression (4) with $b=1$ is realized experimentally is rare. In particular, it was demonstrated in [24] that the experimental results for the textured silicon-based SCs and textured HIT elements near the absorption edge can be described theoretically with a transformed expression of the form (6) with $b > 1$.

Strictly speaking, the realization of the expression (4) or (6) is related to the non-perpendicular transmission of light into the semiconductor both in the case of organic-inorganic perovskites and textured silicon structures. Therefore, it would be more precise to say that the perovskite films synthesized are optically corrugated structures.

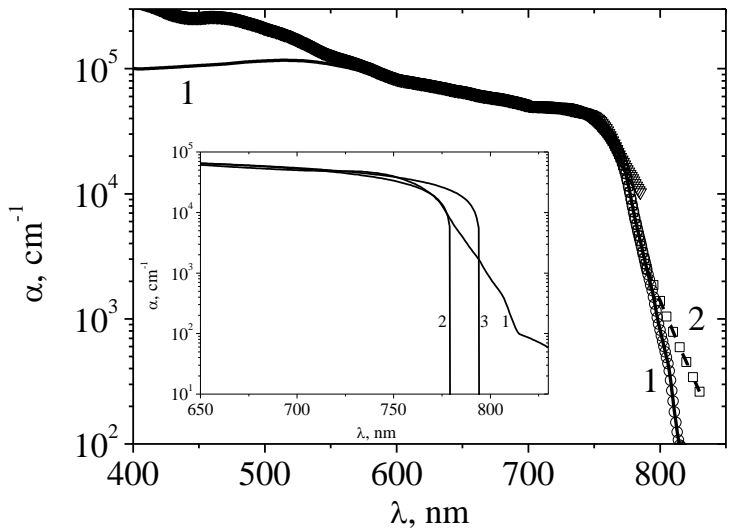

Fig. 4. Spectral curves of the absorption coefficient in $CH_3NH_3PbI_3$. Triangles: experimental data from [22]. Curve 1 is obtained from the solution of the inverse problem; curve 2 – Urbach shift. Circles are obtained using the data from [22] and the solution of the inverse problem. Rectangles are obtained using the data from [22], the solution of the inverse problem, and with Urbach effect taken into account. The inset shows the solution of the inverse problem (curve 1) and the results of Eq. (8) with $E_g = 1.59$ eV (curve 2) and 1.56 eV (curve 3).

Fig. 4 shows the experimental spectral dependence of the absorption coefficient $\alpha(\lambda)$ for the perovskite $CH_3NH_3PbI_3$ obtained in [25]. Unfortunately, the spectral dependence $\alpha(\lambda)$ for the $CH_3NH_3PbI_3$ perovskite from [25] was measured in a restricted range near the absorption edge at $\lambda \leq 800$ nm (triangles). In the wavelength range above 800 nm $\alpha(\lambda)$ can be determined by solving the inverse problem, using the experimental external quantum efficiency of a perovskite $CH_3NH_3PbI_3$ with the thickness d = 100 nm from Ref. [26] and the expression for the absorption coefficient $\alpha = -\ln(1-EQE)/2d$. The values of $\alpha(\lambda)$ found from the solution of the inverse problem together with the experimental curves from [25] are likewise shown in Fig. 4 (curve 1). As seen in the figure, for wavelengths $\lambda \geq 800$ nm, the values from [25] and found from the solution of the inverse problem agree well with each other. The so corrected relation is described by the curve represented by the circles in Fig. 4.

Near the absorption edge, the expression (5) can be Taylor-expanded to become directly proportional to the absorption coefficient $\alpha$ under the condition $4n_r^2 \alpha d / b < 1$. It is well known that the absorption edge in the structurally imperfect films is described by Urbach rule. The empirical absorption coefficient depends on the photon energy in this wavelength range as

$$\alpha_{ur} = \alpha_{ur0} \exp(E_{ph}/E_0), \quad (7)$$

where $\alpha_{ur0}$ is the initial absorption coefficient, $E_{ph}$ is photon energy, and $E_0$ is the characteristic energy, which is of the order of a few tens of meV for not too big deviations from the perfect films [27].

Having analyzed the $V_{ph}$ values near the absorption edge, we established that for the films investigated, Urbach parameters are $\alpha_{ur0} = 2.5 \cdot 10^3$ cm$^{-1}$ and $E_0 = 27$ meV. Using the corrected $\alpha(\lambda)$ dependence (Fig. 4, curve 2), we obtain a curve (see squares), which we will use in the theoretical calculations.

Because organic-inorganic perovskites are direct-bandgap semiconductors, the absorption coefficient $\alpha(\lambda)$ near the absorption edge is given by the formula [28]

$$\alpha(\lambda) = B \frac{\left(\frac{1240}{\lambda} - E_g\right)^{1/2}}{\frac{1240}{\lambda}}, \quad (8)$$

where $B$ is a constant, $E_g$ is the bandgap value, and $\lambda$ is the wavelength in nm.

The inset in Fig. 4 shows the $\alpha(\lambda)$ curve obtained from the solution of the inverse problem (curve 1), and using the expression (8) for the bandgap value of 1.59 eV (curve 2) and 1.56 eV (curve 3). From this figure, one can conclude that the bandgap in the film investigated is 1.59 eV.

Using the expressions (1), (2), (3), and (6) for the $α(λ)$ curves from Fig. 4 allows to build the low-signal surface photovoltage spectral curves and to determine the key parameters (namely, the perovskite film thickness $d$ and the parameter $b$), at which they agree with the experiment near the absorption edge.

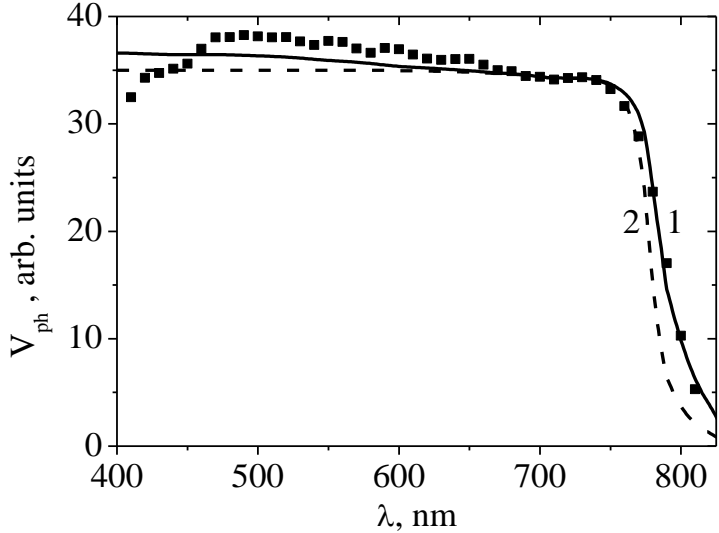

Fig. 5. Spectral curves of $V_{ph}(λ)$ for a film with an ITO layer. Dark circles are the experimental data, curve 1 was obtained theoretically using Eqs. (1) and (6), curve 2 was obtained using Eqs. (1) and (3). The parameters used: $d = 400$ nm, $b = 4$.

Shown in Fig. 5 are the experimental and theoretical values of $V_{ph}(λ)$ for a film with an ITO layer, obtained using the expression (6) and the generalized value of $α(λ)$ (curve 1). As seen from Fig. 5, the experimental $V_{ph}(λ)$-values near the absorption edge agree with the theoretical curve 1 very well if we set $d = 400$ nm and $b = 4$.

If one uses the expression (2), the theory cannot fit the experiment for the film studied (curve 2).

Thus, it follows from this analysis that the film studied here captures light efficiently due to its natural texturing. This conclusion agrees with the microstructure data, see Fig. 3.

Let us note that the spectral curves of the low-signal surface photovoltage for film illumination from the opposite sides are close to each other. This provides an additional evidence of the high diffusion length (exceeding the film thickness), which, according to [29], are of the order of 1 μm in the samples with chlorine added.

Let us now analyse the transmission spectra.

Shown in Fig. 6 are the transmission spectra for a perovskite film without ITO. The respective theoretical formulae are

$$T_{nt} = \frac{(1-R)^2 \exp(-\alpha d)}{1 - R^2 \exp(-2\alpha d)}, \quad (9)$$

and

$$T_t = (1-R)\left(1 + (\alpha(\lambda) l_{ph}(d,b))\right)^{-1}. \quad (10)$$

The expression (9) should be applied for a plane-parallel structure, and (10) is valid for a textured structure (with the coefficient $b$ taken into account).

The transmission spectra from Fig. 6 indicates quite a strong non-uniformity of the thicknesses of the films studied, which leads to a bigger role of the background. Therefore, in order to obtain an agreement between the theoretical and experimental transmission coefficient spectra, a constant background offset had to be subtracted from the initial experimental value of $T$. Let us determine the transmission coefficient according to (10) for a film without the ITO layer using the generalized value for $\alpha(\lambda)$ for the perovskite $CH_3NH_3PbI_{2.98}Cl_{0.02}$ and the values $d = 400$ nm and $b = 4$ (curve 1). In this case, the agreement between experiment and theory is quite good in the wavelength range from 500 nm to 800 nm. However, determination of the transmission coefficient according to (9) gives no agreement between the theoretical curve 2 and the experiment in the whole spectral range.

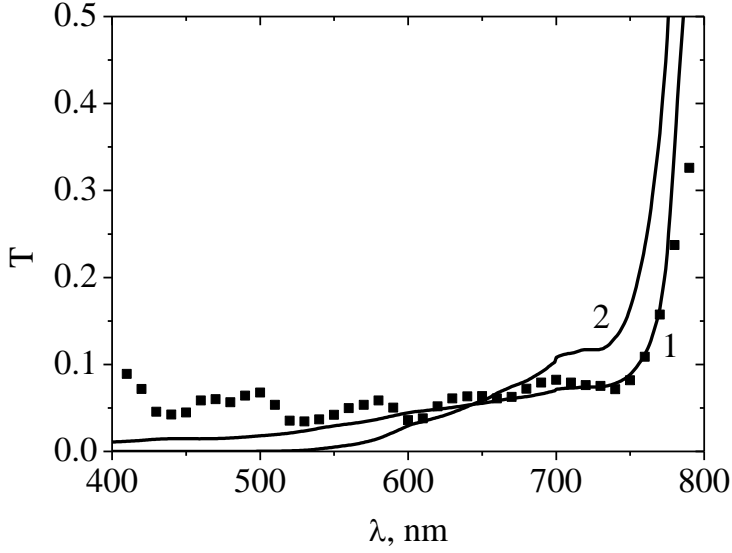

Fig. 6. Transmission spectra of a film without an ITO layer. Dark circles are the experimental data. The theoretical curve 1 was obtained using Eq. (9), the theoretical curve 2 was obtained with Eq. (8). The parameters used: $d = 400$ nm, $b = 4$.

Thus, the experimental transmission spectra of the film without the ITO layer also confirm that the film is naturally textured.

Let us now consider the reasons of the *EQE(λ)* decrease in the short-wavelength range in the perovskite SCs and the spectral dependence of the low-signal surface photovoltage $V_{ph}(\lambda)$, see Fig. 7. There are several reasons for this. For the *EQE*(λ) curves, the short-wavelength decrease is related to light absorption by the ITO film. As for the short-wavelength decrease of $V_{ph}(\lambda)$, it is shown [18] that this decrease has to do with the formation of a layer near the perovskite $CH_3NH_3PbI_{2.98}Cl_{0.02}$ surface with the carrier lifetime shorter than in the bulk [18]. Shown in Fig. 7 are the experimental *EQE*(λ) curves of a SC based on the $FAPbI_3$ perovskite taken from [27] and normalized to the highest value of $V_{ph}(\lambda)$ for a perovskite with *ITO* from this work. The theoretical *IQE*(λ) curve for a $FAPbI_3$-based SC was obtained using the expressions (4) and (6). It agrees with the experiment with *d* taken to be 560 nm. It was established in [23] that in the FAPbI$_3$-based SCs, the perovskite film is naturally textured, and external quantum efficiency absorption edge agrees with Eq. (4) (curve 1). It is shown [20] that this agreement is present for $\alpha d \ll 1$. A comparison between the experimental and theoretical curves shows that the agreement is achieved if one uses the expression (4) at EQE < 0.4 (curve 1). At the same time, the agreement between experiment and the transformed expression (6) with b set to 3 is found not near the absorption edge, but also for EQE reaching the value of ~ 0.9 (curve 2).

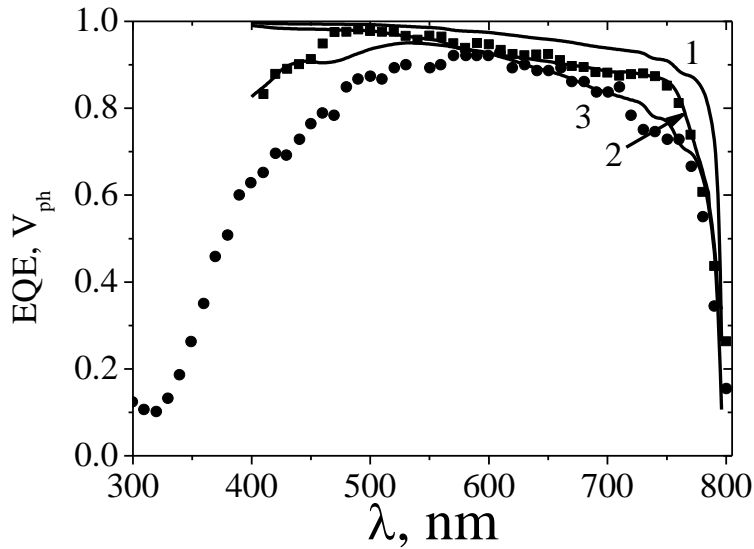

Fig. 7. Experimental (dark and light circles) and theoretical (curves 1 and 2) curves of EQE of a SC based on FAPbI$_3$. Curve 3: $V_f(\lambda)$ of a CH$_3$NH$_3$PbI$_{2.98}$Cl$_{0.02}$ film. The parameters used are as follows. Curve 1: $d = 560$ nm, $b = 1$; curve 2: $d = 560$ nm, $b = 3$; curve 3 - $D_1 = 2.5 \cdot 10^{-2}$ cm$^2$/s, $L_1 = 2.4 \cdot 10^{-6}$ cm, $S_0 = 10^6$ cm/s, $d_1 = 100$ nm, $d = 400$ nm, $b = 4$.

Calculation of the curve $V_{ph}(\lambda)$ for a CH$_3$NH$_3$PbI$_{2.98}$Cl$_{0.02}$ perovskite film, which gives a decrease at short wavelengths, was performed using the following formula for the effective surface recombination velocity

$$S_{eff}(\lambda) = \frac{D_1}{L_1} \frac{S_0 \frac{L_1}{D_1} ch(\frac{1}{\alpha(\lambda) L_1}) + sh(\frac{1}{\alpha(\lambda) L_1})}{S_0 \frac{L_1}{D_1} sh(\frac{1}{\alpha(\lambda) L_1}) + ch(\frac{1}{\alpha(\lambda) L_1})} \quad (11)$$

taken from Ref. [18].

Here, $D_1$ and $L_1$ are, respectively, the diffusion coefficient and diffusion length in the surface layer of thickness $d_1$, and $S_0$ is the effective surface recombination velocity under the condition $\alpha d_1 >> 1$.

The theoretical $V_{ph}(\lambda)$ curve was built using the following parameter values: $D_1 = 2.5 \cdot 10^{-2}$ cm$^2$/s, $L_1 = 2.4 \cdot 10^{-6}$ cm, $S_0 = 10^6$ cm/s, $d_1 = 100$ nm, $d = 400$ nm, $b = 4$ (curve 3).

As seen from Fig. 7, the agreement between $EQE(\lambda)$ and $IQE(\lambda)$ for a FAPbI$_3$-based SC near the absorption edge is good. With respect to the agreement between the experimental $V_{ph}(\lambda)$ curve for the CH$_3$NH$_3$PbI$_{2.98}$Cl$_{0.02}$ film and the theoretical counterpart, it can be considered as satisfactory.

It is interesting to note that the long-wavelength edge of the $EQE(\lambda)$ curve for the FAPbI$_3$-based SC and of the $V_{ph}(\lambda)$ curve for the CH$_3$NH$_3$PbI$_{2.98}$Cl$_{0.02}$ perovskite film are very close to each other.

## Conclusions

In the work, the efficiency of the method of surface photovoltage spectral dependence was demonstrated together with the measurements of the optical properties (transmission and reflection) for the characterization of perovskite films. In particular, the methodology proposed allows one to consistently determine the bandgap value of a perovskite film, as well as its minority carriers diffusion length in the case $L \leq d/3$ or to estimate its value in the opposite case.

As a detailed analysis of the absorption coefficient has shown, in the films mentioned above, Urbach effect plays an essential role near the absorption edge. One can obtain good agreement of the theory with experiment and to deduce an accurate value of the perovskite film bandgap of 1.59 eV only by taking this effect into account.

It has been established by comparing of the experimental spectral curves of the low-signal surface photovoltage with the theory for textured structures that the films studied are naturally textured. This agreement between the theory and the experiment means that the diffusion length of the non-equilibrium charge carriers in the $CH_3NH_3PbI_{2.98}Cl_{0.02}$ perovskite films exceeds the film thickness. This is also supported by the results of a comparison between the spectra of the low-signal surface photovoltage obtained when the film was illuminated from the opposite sides.

It is established that the use of a transformed formula (6) allows one to make the theory consistent with the experiment not only near the absorption edge, but also to the EQE values up to ~0.9.